\documentclass[letterpaper,11pt]{article}
\pdfoutput=1 

\usepackage{jheppub} 

\newcommand{\Tr}{{\rm Tr}}

\title{\boldmath $1/N$ and Loop Corrections in Higher Spin AdS$_4$/CFT$_3$ Duality}

\author[a]{Antal Jevicki,}
\author[b]{Kewang Jin,}
\author[a]{Junggi Yoon}

\affiliation[a]{Department of Physics, Brown University,\\Providence, RI 02912, USA}
\affiliation[b]{Department of Physics, University of Illinois,\\Urbana-Champaign, IL 61801, USA}

\emailAdd{antal\_jevicki@brown.edu}
\emailAdd{kjin@illinois.edu}
\emailAdd{jung-gi\_yoon@brown.edu}

\preprint{{\tt BROWN-HET-1653}}

\abstract{We consider the question of loop corrections (i.e. $1/N$) in the vector model/higher spin duality following the recent work of Giombi and Klebanov \cite{Giombi:2013fka}. The purpose of this paper is to gain further more precise comparison between the two sides of the duality. For CFTs given by 3d $O(N)$ or $U(N)$ vector models we evaluate the leading and one loop partition functions in a variety of geometries. Our calculations are performed in the scheme of collective field theory which was seen in earlier studies to represent a bulk description of Vasiliev higher spin theory. The calculations presented provide data for comparison of small fluctuation determinants giving further evidence for the one-to-one bulk identification between the bi-local and the AdS picture. They also offer insight into the identification of  coupling constants $G$ and $1/N$ of the two descriptions for models based on $O(N)$ symmetry.}

\begin{document} 
\maketitle
\flushbottom

\section{Introduction}
\label{sec:intro}

Recently, vector-like models with $O(N)$ and $U(N)$ symmetries at their critical points were seen to exhibit duality \cite{Klebanov:2002ja, Sezgin:2002rt, Giombi:2009wh} with higher spin gravitational theories of Vasiliev \cite{Vasiliev:1990en, Vasiliev:1992av, Bekaert:2005vh}. Typically in 3d vector field theory, there are two conformally invariant fixed points, the free UV fixed point and the interacting IR fixed point. The higher spin duals to these two fixed points are given by the same Vasiliev theory but with different boundary conditions in the quantization of the bulk scalar field. This Vector Model/Higher Spin correspondence was also extended to the supersymmetric case \cite{Leigh:2003gk, Sezgin:2003pt}, Chern-Simons theories \cite{Aharony:2011jz, Giombi:2011kc} and de Sitter space \cite{Anninos:2011ui, Das:2012dt}. Furthermore one also has the very rich and nontrivial lower dimensional dualities involving 2d Minimal Model CFTs and 3d Higher Spin Gravities \cite{Gaberdiel:2010pz, Chang:2011mz, Jevicki:2013kma}. All these dualities have received definite support based on evaluation of three-point correlation functions, finite temperature partition functions and study higher conservation laws.

In a series of papers \cite{Das:2003vw, Koch:2010cy, Jevicki:2011ss} an explicit operator construction of the (dual) Higher Spin AdS theory in terms of collective fields was developed. This approach provides a framework for a one-to-one reconstruction of AdS spacetime, higher spin fields in the bulk and their $1/N$ interactions. Higher order calculations that were performed concerned the 1-loop correction to the free energy \cite{Das:2003vw}, correlation functions \cite{deMelloKoch:1996mj}, and an investigation of the (non)triviality of the theory \cite{deMelloKoch:2012vc} based on free fields.

The purpose of this paper is to study further the question of loop corrections (i.e. $1/N$) in the higher spin duality. We follow up the earlier work of \cite{Das:2003vw} and the recent work of Giombi and Klebanov \cite{Giombi:2013fka}. These calculations concern the evaluation of partition functions at one loop in the collective and also in the AdS version of the theory. In both cases, the one-loop corrections follow from the quadratic Laplacians
\begin{equation}
\Tr\log\Box_{\text{bi-local}} \qquad \text{and} \qquad 
\Tr\log\frac{\Box_{\text{hs}}}{\Box_{\text{gh}}} \ .
\end{equation}
In the light-cone gauge the Laplacians can be shown to be equal
\begin{equation}
\Box_{\text{bi-local}}=2\partial_+\partial_--\left(\frac{p_1^2}{p_1^+}+\frac{p_2^2}{p_2^+}\right)\left(p_1^++p_2^+\right)=2\partial_+\partial_--\left(\partial_x^2+\partial_z^2\right)=\nabla_{\text{hs}}
\end{equation}
as a consequence of the spacetime mapping established in \cite{Koch:2010cy}.\footnote{A very similar identification of AdS space in light-cone QCD was found in \cite{Brodsky:2006uqa, Brodsky:2013dca}.} Due to gauge invariance one could then expect identical results for their one-loop contributions in general. However, since one considers backgrounds which do not always easily fit into the light-cone gauge, explicit calculations are nevertheless worthwhile. They also serve as the purpose for understanding more completely the nature of loop corrections in higher spin duality. In particular in the heat-kernel AdS calculation the suggestion was made in \cite{Giombi:2013fka} that the identification of the gravitational coupling constant should be taken as $G=1/(N-1)$ for the dualities based on the $O(N)$ symmetry group (no such change was found for the $U(N)$ case). Our results shed some light on this identification. First of all collective theory shows that in addition to the determinant there is one further contribution of $O(1)$ associated with the measure appearing in the functional integration. The measure does provide the needed cancellation at one loop (as noticed originally in \cite{Das:2003vw}) allowing the standard identification of $G=1/N$. However collective field theory also indicates a freedom of a finite (re)normalization of $G$ into $1/(N-1)$ as we discuss in the text. These two expansion schemes are compatible, as one can re-expand results of one into another.

The content of this paper goes as follows. In section \ref{sec:collective} we consider first the finite temperature case of the CFT reviewing an earlier work of \cite{Das:2003vw}. This example already contains some of the basic effects that will be observable in the rest of the calculations. We then present details of the bi-local calculation in the case of $S^3$ (the example of \cite{Giombi:2013fka, Klebanov:2011gs}) and point out the role of the measure. In section \ref{sec:thermal} we proceed to the other phase of the theory discussing the evaluation of the partition function in thermal AdS both by the heat-kernel method and in the bi-local collective field framework. Some conclusions are given in section \ref{sec:conclusions}.

\section{Collective approach to the loop corrections}
\label{sec:collective}

The collective theory describes the large $N$ dynamics of bi-local collective fields. These fields have the property that they close under the Schwinger-Dyson equations. They represent a more general set than the conformal currents and contain an additional dimension. As such they are natural candidates for representing the bulk AdS$_4$ theory. This is supported by the fact that an effective, collective field action with the property that the associated functional integral exactly evaluates the $O(N)$ singlet partition function and correlation functions of bi-local operators. The diagramatics accomplished by this reformulation is that of Witten diagrams.

The exact partition function of the free vector model with $N$ components in terms of the bi-local field $\Phi\left(x,y\right)$ is given by \cite{Das:2003vw}
\begin{equation}
Z=\int \mathcal{D}\Phi\left(x,y\right) J(x,y) e^{-S\left[\Phi\left(x,y\right)\right]}
=\int \mathcal{D}\Phi\left(x,y\right) \mu \, e^{-S_{\text{col}}\left[\Phi\left(x,y\right)\right]}
\label{partition}
\end{equation}
where $J(x,y)$ is the Jacobian (generated from the change of variables from the fundamental vector fields to the bi-local fields), and the collective action reads
\begin{equation}
S_{\text{col}}=N \int d^3x \left(-\left.\Delta_x\Phi\left(x,y\right)\right|_{x=y}\right)-\frac{1}{2} N \Tr\log \Phi \ .
\label{collaction}
\end{equation}
The (integration) measure $\mu$ in (\ref{partition}) is computed to be
\begin{equation}
\mu=\left(\det\Phi\right)^{-\kappa}
\end{equation}
where the power $\kappa$ depends on the underlying symmetry of the vector fields. For the $O(N)$ case, the bi-local field $\Phi\left(x,y\right)=\frac{1}{N} \vec{\phi}\left(x\right)\cdot\vec{\phi}\left(y\right)$ is symmetric and $\kappa=\frac{1}{2}\left(K+1\right)$ with $K=\sum_{k}1$ the `volume' of the momentum space. While for the $U\left(N\right)$ case, the bi-local field $\Phi\left(x,y\right)=\frac{1}{N}\vec{\phi}^*(x)\cdot\vec{\phi}\left(y\right)$ is Hermitian and we have $\kappa=K$. The details of this derivation can be found in the appendix \ref{app:measure}. In the Riemann zeta-function regularization (employed in \cite{Giombi:2013fka} and also here), we have set $K=0$ so that the measure is simplified to be
\begin{equation}
\mu = \begin{cases}
1 & \text{for}~ U(N) \\
(\det \Phi)^{-1/2} & \text{for}~ O(N)
\end{cases} \ .
\end{equation}

We mention that the action on this representation scales with $N$ and the interactions generated are consequently given in powers of $1/N$ and the measure would contribute in the subleading orders. It is also relevant to point out at the outset that the measure in this collective representation leads to a contribution of the same form as the $\Tr\log$ term in the action (\ref{collaction}) (which sets the coupling constant). Consequently one can equivalently include the measure term into the action obtaining an effective coupling constant. We will return to this issue of interpretation in section \ref{sec:interpretation} after presenting the one loop calculations.

\subsection{Partition function on $S^1 \times \mathbb{R}^2$}

We start by reviewing first the one-loop calculation performed in \cite{Das:2003vw} for the $S^1 \times \mathbb{R}^2$ partition function. This case already demonstrates some of the features of the one loop determinant that will be general and central to the issues raised in the Introduction.

One develops the expansion as usual by shifting the background bi-local field
\begin{equation}
\Phi\left(x_1,x_2\right)=\Phi_0\left(x_1,x_2\right)+\frac{1}{\sqrt{N}}\eta\left(x_1,x_2\right)
\end{equation}
where $\Phi_0\left(x_1,x_2\right)$ represents the stationary point of the collective action (\ref{collaction}). In the momentum space representation one has the Fourier transformed field $\tilde{\Phi}_0(k_1,k_2)$ with the momenta $k_{1,2}=(\nu_n,\vec{k}_{1,2})$ where the Matsubara frequency is given by $\nu_n=\frac{2 \pi n}{\beta}$ and $\beta$ as the inverse of temperature $T$.

The zeroth-order collective action is now given by
\begin{equation}
S_{\text{col}}^{\left(0\right)}=N \sum_{k} k^2 \tilde{\Phi}_{0}(k,-k)-\frac{1}{2} N \Tr\log \tilde{\Phi}_0 \ .
\end{equation}
Translation invariance implies $\tilde{\Phi}_{0}(k_1,k_2)=\xi\left(k_1\right)\delta_{k_1,-k_2}$, and we get
\begin{equation}
S_{\text{col}}^{(0)}=N\sum_{k} k^2 \xi\left(k\right) -\frac{1}{2}N \sum_{k}\log\left[\xi\left(k\right)\right] \ .
\end{equation}
By the saddle point method one determines $\xi\left(k\right)= \frac{1}{2k^2}$, and the background field is
\begin{equation}
\Phi_0\left(x,y\right)=\frac{1}{\left(2\pi\right)^2\beta}\sum_n\int d^2\vec{k} \frac{1}{2\left(\vec{k}^2+\left(\frac{2\pi n}{\beta}\right)^2\right)}e^{ik\cdot\left(x-y\right)}
\end{equation}
which is nothing but the free two point function $\langle \phi^i (x) \phi^i (y)\rangle$ of the bi-local operators. Evaluating the action at the background value produces the leading contribution to the free energy 
\begin{equation}
F^{\left(0\right)}=S_{\text{col}}^{(0)}=\frac{N}{2}\sum_n\sum_{\vec{k}}\log \left[\vec{k}^2+\left(\frac{2\pi n}{\beta}\right)^2\right]
\end{equation}
which is precisely the free energy of $N$ free bosons $F^{\left(0\right)}=\frac{N}{2} \Tr\log \partial^2$. At high temperature, the free energy scales as $F^{(0)} \sim - N \zeta(3) T^2$, producing the lower phase of \cite{Shenker:2011zf}.

To evaluate the 1-loop contribution, one expands the collective action $S_{\text{col}}$ to the quadratic order in the fluctuations $\eta$ :
\begin{eqnarray}
S_{\text{col}}^{\left(2\right)}&=&\frac{1}{4}\Tr\left(\eta\Phi_0^{-1}\eta\Phi_0^{-1}\right)\equiv\Tr\left(\eta\Box\eta\right)\cr
&=&2 \sum_{k_1>k_2} k_1^2k_2^2\eta_{k_1,-k_2}\eta_{k_2,-k_1}+\sum_{k}\left(k^2\right)^2\eta_{k,-k}\eta_{k,-k} \ .
\end{eqnarray}
Then the one-loop free energy comes as the determinant of the generalized (bi-local) Laplacian. Because of the product form the determinant factorizes and one obtains
\begin{eqnarray}
F^{\left(1\right)}&=&\frac{1}{2}\Tr\log\left(\Box\right)=\sum_{k_1>k_2}\frac{1}{2}\log\left(k_1^2k_2^2\right)+\sum_{k}\log (k^2)\cr
&=&\frac{1}{2}\left(K+1\right)\sum_{k}\log (k^2)
\label{oneloop}
\end{eqnarray}
with a surprising finding that the bi-local determinant produces the local field contribution (with a factor $K+1$). This pre-factor is most significant as it is associated with the counting of bi-local degrees of freedom. With a zeta-function regularization the infinite `volume' $K$ would be set to $0$ and the result corresponds to the $N=1$ single field expression. This is a prototype of the result that was also observed in \cite{Giombi:2013fka}, namely the evaluation of the AdS higher spin determinant in the heat-kernel method using the zeta-function regularization gave the $N=1$ CFT result.

The collective representation however contains one other contribution of order $\mathcal{O}\left(N^0\right)$. It comes from the measure $\mu$ evaluated at the stationary point
\begin{equation}
\Delta F^{\left(1\right)}=\frac{1}{2}\left(K+1\right)\Tr\log \Phi_0=-\frac{1}{2}\left(K+1\right)\sum_{k}\log\left(k^2\right) \ .
\end{equation}
Thus the total one-loop correction to the free energy is found to be
\begin{equation}
F^{\left(1\right)}_{\text{total}}=F^{\left(1\right)}+\Delta F^{\left(1\right)}=0 \ .
\end{equation}
This complete cancellation between the determinant and the measure contribution therefore assures the required result $0$.

To recapitulate, the one loop determinant of fluctuations produces an answer identical to that of $N$ free scalars in $d=3$ but with $N$ replaced by $K+1$. If $K$ (which is infinite) is set to $0$ by regularization the result then corresponds to $N=1$, i.e. to that of a single scalar field. This is what was also found in \cite{Giombi:2013fka} and will be the case in all the other examples that follow. One can trace its origin of this to the bi-local nature of degrees of freedom in this theory. In particular the appearance of $K+1$ in $O(N)$ theories (and $K$ in $U(N)$ theories) is associated with the fact that the fields can be encoded into a symmetric matrix appearing naturally in the bi-local description. Equally importantly in the collective higher spin representation, one also has a measure in the functional integral which leads to cancellation and the result $F_{\rm total}^{(1)}=0$ at one loop.

\subsection{Partition function on $S^3$}

We now consider the partition function on $S^3$, the example that was considered in \cite{Giombi:2013fka}. One follows the same procedure described in detail as in the previous section, the only difference being the explicit expressions for the eigenfunctions and eigenvalues. 

Using spherical harmonics of $S^3$, the Fourier transformation of the bi-local field is 
\begin{align}
\Phi\left(x_1,x_2\right)=\sum_{\vec{k}_1,\vec{k}_2}\Phi_{\vec{k}_1,\vec{k}_2}Y_{\vec{k}_1}\left(x_1\right)Y_{\vec{k}_2}\left(x_2\right)
\end{align}
where $\vec{k}$ denotes a full set of quantum numbers $\vec{k} \equiv (l,n,m)$ and
\begin{eqnarray}
l&=&0,1,2,\cdots,\cr
n&=&0,1,2,\cdots,l \cr
m&=&-n,-\left(n-1\right),\cdots, n-1,n \ . \nonumber
\end{eqnarray}
Denoting the conjugate label of $\vec{k}$ as $\vec{k}^* \equiv (l,n,-m)$, the classical background field is now
\begin{align}
\Phi_0\left(x_1,x_2\right)=\sum_{\vec{k}}\frac{\left(-1\right)^m}{2\lambda(\vec{k})}Y_{\vec{k}}\left(x_1\right)Y_{\vec{k}^*}\left(x_2\right)
\end{align}
where $\lambda(\vec{k})$ are the eigenvalues of the Laplacian on $S^3$ :
\begin{align}
\lambda(\vec{k})=\left(l+\frac{3}{2}\right)\left(l+\frac{1}{2}\right) \ .
\end{align}
From the background field, one can calculate the leading free energy
\begin{align}
F^{\left(0\right)}=S_{\text{col}}^{(0)}\left(\Phi_0\right)=\frac{N}{2}\sum_{\vec{k}}\log \lambda(\vec{k})=N\left(\frac{1}{8}\log2-\frac{3\zeta\left(3\right)}{16\pi^2}\right) \ ,
\end{align}
where we have used the Riemann zeta function regularization as in \cite{Klebanov:2011gs}.

For the one-loop contribution, following the same procedure which leads to (\ref{oneloop}), we have the result
\begin{align}
F^{\left(1\right)}=\frac{1}{2}\left(K+1\right)\sum_{\vec{k}}\log \lambda(\vec{k}) \ .
\end{align}
In the zeta function regularization, the constant $K$ gives
\begin{align}
K=\sum_{\vec{k}}1=\sum_{l=0}^\infty\sum_{n=0}^l\sum_{m=-n}^n1=\zeta\left(-2\right)=0 \ .
\end{align}
Therefore, the one-loop contribution to free energy is
\begin{align}
F^{\left(1\right)}=\frac{1}{8}\log2-\frac{3\zeta\left(3\right)}{16\pi^2}
\end{align}
which is exactly the contribution from a single scalar field. Notice that a bi-local field in the $U\left(N\right)$ vector model is not symmetric, but Hermitian, the one-loop free energy of $U\left(N\right)$ is $F^{\left(1\right)}_{U\left(N\right)}=K\sum_{\vec{k}}\log\lambda(\vec{k})$. After regularization, the free energy of $U\left(N\right)$ vector model vanishes $F^{\left(1\right)}_{U\left(N\right)}=0$ as a result of $K=0$. This also agrees with \cite{Giombi:2013fka}.

Remember there is another correction to the one-loop free energy from the measure $\mu$ by plugging in the background bi-local field
\begin{align}
\Delta F^{\left(1\right)}=\frac{1}{2}\left(K+1\right)\Tr\log \Phi_0=-\frac{1}{2}\left(K+1\right)\sum_{\vec{k}}\log  \lambda(\vec{k}) \ .
\end{align}
The total one-loop free energy is therefore
\begin{align}
F^{\left(1\right)}_{\text{total}}=F^{\left(1\right)}+\Delta F^{\left(1\right)}=0 \ .
\end{align}
The cancellation of one-loop free energy by the contribution of the measure also occurs in the case of $U\left(N\right)$.

\subsection{Interpretation of the results}
\label{sec:interpretation}

Collective higher spin field theory based on  bi-local fields realizes AdS/CFT duality in the bulk through the path integral
\begin{align}
Z=\int d\Phi\left(x,y\right)\mu\left[\Phi\right]e^{-S_{\text{col}}[\Phi]}=Z\left(G=\frac{1}{N}\right)
\end{align}
where  the  action is given by
\begin{align}
S_{\text{col}}=S_0 - \frac{N}{2} \Tr\log \Phi \ .
\label{trlog}
\end{align}
Compared with the original CFT action $S_0$, we have an extra $\mathcal{O}(N)$ term given by the $\Tr\log$ term in (\ref{trlog}) responsible for the $G=1/N$ expansion, and a $\mathcal{O}\left(N^0\right)$ measure term
\begin{align}
\mu=\left(\det \Phi\left(x,y\right)\right)^{-\kappa}
\end{align}
with
\begin{align}
\kappa=\begin{cases}
K& \qquad \text{for}~ U\left(N\right) \\
\frac{1}{2}\left(K+1\right)&\qquad \text{for}~ O\left(N\right)\\
\end{cases} \ .
\end{align}
These two terms both represent the quantum effects, they specifically come from the Jacobian arising in the change of variables from $N$-component scalar fields $\phi^i(x)$ to the singlet bi-local fields $\Phi(x,y)=\phi^i(x)\phi^i(y)$ :
\begin{align}
\log J=\frac{1}{2}\left(N-2\kappa\right)\Tr\log \Phi \ .
\end{align}
Altogether the action (expandable in $1/N$) and the measure of lower order define the systematic $1/N$ expansion of the theory.

But the collective field representation offers another possibility. One notices the fact the measure term and the additional term contributing to the action have the same functional form. This then allows an alternative splitting for example with the whole $\log J$ added to the action 
\begin{align}
Z=\int d\Phi\left(x,y\right) e^{\partial^2\Phi+\frac{1}{2}\left(N-2\kappa\right)\Tr\log\Phi}=Z\left(G^*\right) \ .
\end{align}
This leads to a formulation without any measure and an effective coupling constant given by
\begin{align}
G^*=\frac{1}{N-2\kappa} \ .
\end{align}
One can be worried about this scheme considering the fact that this represents an infinite renormalization of the coupling constant. But in the case of $O\left(N\right)$ models where $2\kappa=K+1$ (and $K$ is infinite), we can include the $2 \kappa=1$ part into the coupling resulting in
\begin{align}
Z=\int d\Phi (x,y) \, \mu' \left[\Phi\right] e^{-\left(N-1\right)S_{\text{col}}}=Z\left(G'=\frac{1}{N-1}\right)
\end{align}
and an expansion based on the new coupling constant
\begin{align}
G'=\frac{1}{N-1} \ .
\end{align}
In this case, the measure is $\mu'=\left(\det \Phi \right)^{-\frac{1}{2}K}$. Employing a regularization which sets $K=0$ we have the expansion parameter  $G'=1/(N-1)$ and no extra measure. This would be in agreement with the identification suggested in \cite{Giombi:2013fka}.

In general, gravitational theories come with a nonzero measure \cite{Faddeev:1973zb}. For example, the functional measure in (quantized) general relativity was computed in \cite{Fradkin:1974df, Kaku:1976xe} to be
\begin{align}
\mu=\prod_x \Bigl[ g^{7/2}(x) g^{00}(x) \prod_{\sigma \le \lambda} dg^{\sigma\lambda}(x) \Bigr] \ ,
\end{align}
where $g \equiv \det g_{\mu\nu}$. It contributes infinite $\delta^{(4)}(0)$ terms in perturbation theory canceling analogous divergences of Feynman diagrams. In dimensional or zeta function regularization, such terms are set to $0$. 

In Vasiliev theory, one has not yet worked out the measure (evaluating it would require the use of an action). But, the existence of a collective representation for this theory would indicate that there will be an analogous measure. If what we have learned in the collective representation is telling, then in a regularization where such a measure is removed, one could define an effective coupling constant so that expansion would naturally become $G'=1/(N-1)$ for $O(N)$ theories as compared to $G=1/N$ for $U(N)$ duals. We mention however that for non-perturbative studies involving the Hilbert space (and entropy) it might not be appropriate to use a regularization which removes the measure. Such is for example the case of dS/CFT \cite{Das:2012dt}. In any case it is of interest to evaluate the one loop measure of higher spin theories.

Another possibility was suggested by Leigh and Petkou \cite{Leigh:2012mz}. On the field theory side, an explicit symmetry breaking from $O(N) \to O(N-1)$ can be triggered by adding a singleton deformation. Such deformation, in the bulk, can be absorbed by the higher spin fields with a shift of the parameter $N \to N+1$. Therefore, the singleton deformation breaks higher-spin symmetry and generates a $1/N$ correction to the free energy.

\section{One loop partition functions in thermal AdS$_4$}
\label{sec:thermal}

We now proceed to the study (and evaluation) of the free energy in the case of another geometry (thermal AdS$_4$). This actually represents a different phase of the theory, involving the phase transition described in \cite{Shenker:2011zf}. In this case we perform calculations both in the AdS heat-kernel version and the bi-local collective version. The purpose is first of all to observe an agreement between the two calculations and also to see that the phenomena put forward in section \ref{sec:collective} persist in the case of a different background. This will happen even though the physics of the two phases (as emphasized in \cite{Shenker:2011zf}) is very different.

\subsection{The Heat Kernel method}

Thermal AdS$_4$ is defined by periodicity conditions on the Euclidean time variable $\tau \in \left[0,\beta\right]$. One expands the metric $g$ around the AdS background which is taken the same (static) solution as the AdS vacuum $g=g_{\text{AdS}}+\eta$. In \cite{Gopakumar:2011qs,Gupta:2012he}, the partition functions of higher spin theories in odd dimensional AdS spaces are explicitly calculated using the heat kernel method. One can follow exactly the same method in performing the calculations in AdS$_4$. The partition function of massless spin-$s$ field is then
\begin{eqnarray}
Z_{\left(s\right)} &=& \exp\left[-\frac{1}{2}\Tr\log\frac{\left(-\nabla^2+s^2-2s-2\right)}{\left(-\nabla^2+s^2-1\right)}\right] \cr
&=&\prod_{m=1}^\infty\left[\frac{\left(1-q^{s+m+1}\right)^{2s-1}}{\left(1-q^{s+m}\right)^{2s+1}}\right]^{\frac{m\left(m+1\right)}{2}}
=\prod_{m=1}^\infty\frac{1}{\left(1-q^{s+m}\right)^{m\left(m+2s\right)}} \ .
\end{eqnarray}
The partition function of the massless scalar field is
\begin{eqnarray}
\log Z_{\left(0\right)}&=&-\frac{1}{2}\log\det\left(-\bigtriangleup+M_{\left(0\right)}^2\right)=\sum_{m=1}^\infty \frac{1}{m\left(1-q^m\right)^{3}}q^{m\left(\frac{3}{2} \pm \sqrt{\frac{9}{4}+M_{\left(0\right)}^2}\right)} \cr
&\equiv&\sum_{m=1}^\infty \frac{q^{m\Delta_{\left(0\right)}}}{m\left(1-q^m\right)^{3}}=\sum_{m=1}^\infty\binom{m+1}{2}\log\frac{1}{1-q^{\Delta_{\left(0\right)}+m-1}}
\end{eqnarray}
resulting in
\begin{align}
Z_{\left(0\right)}=\prod_{m=1}^\infty \frac{1}{\left(1-q^{\Delta_{\left(0\right)}+m-1}\right)^{\frac{m\left(m+1\right)}{2}}}
\end{align}
where $\Delta_{(0)}$ is the scaling dimension of the bulk scalar field.

For the UV fixed point, which corresponds to $\Delta_{(0)}=1$, we have the partition function for the scalar field as
\begin{equation}
Z_{\left(0\right)}=\prod_{m=1}^\infty \frac{1}{\left(1-q^m\right)^{\frac{m\left(m+1\right)}{2}}} \ .
\end{equation}
Multiplying with all the higher spin contributions, the total one loop partition function of higher spin gravity (which corresponds to the $U(N)$ vector model on the boundary) is
\begin{align}
Z=\prod_{s=0}^\infty Z_{\left(s\right)}=\frac{1}{\left(1-q\right)\left(1-q^2\right)^3}\prod_{m=2}^\infty\frac{1}{\left(1-q^m\right)^{\binom{m+1}{2}}\left(1-q^{m+1}\right)^{3\binom{m+1}{2}+4\binom{m+1}{3}}} \ .
\end{align}
Therefore, the associated free energy is
\begin{align}
F=-\log Z =\sum_{m=1}^\infty\left(\frac{2}{3}m^3+\frac{1}{3}m\right)\log\left(1-q^m\right)=-\sum_{k=1}^\infty\frac{1}{k}\frac{q^k\left(1+q^k\right)^2}{\left(1-q^k\right)^4}
\label{un}
\end{align}
which agrees with eq. (10) of \cite{Shenker:2011zf}.

Also, for the minimal higher spin theory which includes only the even higher spin fields, the free energy is then
\begin{align}
\begin{split}
F_{\rm min}=&-\log Z_{\rm min}=\sum_{m=1}^\infty\frac{m\left(m+1\right)}{2}\log\left(1-q^m\right)+\sum_{m,s=1}^\infty m\left(m+4s\right)\log\left(1-q^{2s+m}\right)\\
=&\sum_{m=1}^\infty\frac{m\left(m+1\right)}{2}\log\left(1-q^m\right)+\sum_{m=3}^\infty\sum_{s=1}^{\left[\frac{m-1}{2}\right]}\left(m^2-4s^2\right)\log\left(1-q^m\right)\\
=&-\sum_{k=1}^\infty\frac{q^k}{k}\frac{\left(1+q^k+4q^{2k}+q^{3k}+q^{4k}\right)}{\left(q^k+1\right)^2\left(q^k-1\right)^4}=-\frac{1}{2}\sum_{k=1}^\infty\frac{q^k}{k}\left[\frac{\left(1+q^k\right)^2}{\left(1-q^k\right)^4}+\frac{1+q^{2k}}{\left(1-q^{2k}\right)^2}\right] \ .
\label{on}
\end{split}
\end{align}
This result will be seen to agree with the singlet $O(N)$ model case using the collective field method.

\subsection{The Hamiltonian method}

For completeness, let us describe how the heat kernel evaluations can be equivalently obtained by a Hamiltonian method as described in \cite{Gibbons:2006ij}. The one-particle partition function of a massless field in AdS$_4$ as a function of the temperature $T=\beta^{-1}$ and the chemical potential $\Omega$ is written as
\begin{align}
Y\left(\beta,\Omega\right)=\sum_{E,j} e^{-(\beta E+\alpha j)} \ .
\end{align}
For the representations which are relevant for the UV fixed point, we have
\begin{align}
Y_{\left(1,0\right)}(\beta,\Omega=0)=&\frac{e^{2\beta}}{\left(e^\beta-1\right)^3}\\
Y_{\left(s+1,s\right)}(\beta,\Omega=0)=&\frac{e^{\left(1-s\right)\beta}\left[\left(2s+1\right)e^\beta+1-2s\right]}{\left(e^\beta-1\right)^3}\qquad (s \ge 1) \ .
\end{align}
From the single-particle partition function, one deduces energy spectrum and the degeneracies
\begin{alignat}{5}
&D\left(1,0\right)\quad &&:\quad E_n=n,\quad&&d_n=\frac{1}{2}n\left(n+1\right),&&\qquad\left(n\ge 1\right)\\
&D\left(s+1,s\right)\quad &&:\quad  E_n=n,\quad&&d_n=n^2-s^2,&&\qquad \left(n\ge s\ge 1\right) \ .
\end{alignat}
Therefore, through the formula
\begin{align}
F=\sum_{n} d_n\log\left(1-e^{-\beta E_n}\right)
\end{align}
one can obtain free energies of massless particles in AdS$_4$ as
\begin{align}
F_{\left(0\right)}=&\sum_{m=1}^\infty\frac{m\left(m+1\right)}{2}\log\left(1-q^{m}\right)\\
F_{\left(s\right)}=&\sum_{m=s}^\infty\left(m^2-s^2\right)\log\left(1-q^{m}\right)\qquad (s \ge 1)
\end{align}
where $q=e^{-\beta}$. Thus, the total free energy is
\begin{align}
\begin{split}
F=&\sum_{s=0}^\infty F_{\left(s\right)}=F_{\left(0\right)}+\sum_{s=1}^\infty\sum_{m=s}^\infty\left(m^2-s^2\right)\log\left(1-q^{m}\right)\\
=&F_{\left(0\right)}+\sum_{m=1}^\infty \sum_{s=1}^{m}\left(m^2-s^2\right)\log\left(1-q^{m}\right)=\sum_{m=1}^\infty \frac{1}{3}m\left(2m^2+1\right)\log\left(1-q^m\right)\\
=&-\sum_{m=1}^\infty \sum_{k=1}^\infty \frac{1}{3}m\left(2m^2+1\right)\frac{q^{mk}}{k}=-\sum_{k=1}^\infty \frac{1}{k}\frac{q^k\left(1+q^k\right)^2}{\left(1-q^k\right)^4}
\end{split}
\end{align}
which agrees with \eqref{un}. Also, one can add up only the even spin fields and the scalar field contributions to get
\begin{align}
\begin{split}
F_{\rm min}=\sum_{s=0}^\infty F_{\left(2s\right)}=&-\frac{1}{2}\sum_{k=1}^\infty\frac{q^k}{k}\left[\frac{\left(1+q^k\right)^2}{\left(1-q^k\right)^4}+\frac{1+q^{2k}}{\left(1-q^{2k}\right)^2}\right]
\end{split}
\end{align}
which agrees with \eqref{on}.

\subsection{The collective field theory approach}

We will now describe the evaluation of the partition function in the bi-local picture. Since the background is given by the ground state solution it is appropriate to use the Hamiltonian (single-time) representation of the bi-local theory \cite{Koch:2010cy}. The full nonlinear collective Hamiltonian for the equal-time bilocal field (and its canonical conjugate) reads
\begin{align}
\begin{split}
H&=\frac{1}{2}\int d\vec{x}d\vec{y}d\vec{z} \, \Pi\left(\vec{x},\vec{y}\right) \Psi\left(\vec{y},\vec{z}\right)\Pi\left(\vec{z},\vec{x}\right)+\frac{1}{2}\int d\vec{x}d\vec{y} \, \Pi\left(\vec{x},\vec{y}\right) \Psi\left(\vec{y},\vec{x}\right)\Pi\left(\vec{x},\vec{x}\right)\\
&+\frac{1}{2}\int d\vec{x}d\vec{y} \, \Pi\left(\vec{x},\vec{x}\right) \Psi\left(\vec{x},\vec{y}\right)\Pi\left(\vec{y},\vec{x}\right)+\frac{1}{2}\int d\vec{x} \, \Pi\left(\vec{x},\vec{x}\right) \Psi\left(\vec{x},\vec{x}\right)\Pi\left(\vec{x},\vec{x}\right)\\
&+\frac{1}{2}\int d\vec{x} \left(-\left.\Box_{\vec{x}} \Psi\left(\vec{x},\vec{y}\right)\right|_{\vec{y}=\vec{x}}\right)+\frac{N^2}{8} \Tr\Psi^{-1}+\Delta V
\label{hamiltonian}
\end{split}
\end{align}
where $\Box_{\vec{x}}$ is the Laplacian on $S^2$ and the counterterms (which are lower orders in $1/N$) are
\begin{align}
\Delta V=\left(-\frac{N}{4}\left(K+1\right)+\frac{1}{8}\left(K+1\right)^2\right)\Tr\Psi^{-1} \ .
\end{align}
The first five (integral) terms on the RHS of (\ref{hamiltonian}) comes from a direct rewriting of the original Hamiltonian (of the vector fields) in terms of the bi-local fields (after a repeated use of the chain rule) (see \cite{Jin:2013lqa} for details). The rest terms in (\ref{hamiltonian}) (including the interaction term $\Tr \Psi^{-1}$ and the counterterm $\Delta V$) arises from a similarity transformation to make the Hamiltonian Hermitian. This is in the same spirit as the Jacobian present in the action approach, and the counterterm is related to the lower order measure.

The collective Hamiltonian (\ref{hamiltonian}) is well suited to perform a $1/N$ expansion after the rescaling $\Psi \to N \Psi$ and $\Pi \to \Pi/N$. By expanding $\Psi\left(\vec{x},\vec{y}\right)$ around the background field $\Psi=\Psi_0+\frac{1}{\sqrt{N}}\eta$, and similarly for the conjugate momenta $\Pi=\sqrt{N}\pi$, one can show that the leading Hamiltonian $H^{\left(0\right)}$ of order $\mathcal{O}\left(N\right)$ is
\begin{align}
\begin{split}
E^{(0)}=&H^{\left(0\right)}=\frac{N}{2}\int d^2x \left[-\left.\nabla_{\vec{x}}^2\Psi_0\left(\vec{x},\vec{y}\right)\right|_{\vec{y}=\vec{x}}\right]+\frac{N}{8}\Tr\Psi_0^{-1} \cr
=&\frac{N}{2}\sum_{\vec{k}}\left(l+\frac{1}{2}\right)=\frac{N}{4}\sum_{l=0}\left(2l+1\right)^2
\end{split}
\end{align}
which is exactly the ground state energy of $N$ free bosons. 

The one-loop calculation follows similarly as the covariant formulation used in the previous section. The quadratic Hamiltonian of order $\mathcal{O}\left(N^0\right)$ is
\begin{align}
H^{\left(2\right)}=\frac{1}{2}\sum_{\vec{k}_1\le \vec{k}_2} \left[\pi_{\vec{k}_1,\vec{k}_2^*}\pi_{\vec{k}_2,\vec{k}_1^*}+\eta_{\vec{k}_1,\vec{k}_2^*}\omega^2_{\vec{k}_1,\vec{k}_2}\eta_{\vec{k}_2,\vec{k}_1^*}\right]
\end{align}
where $\vec{k}=\left(l,m\right)$ and $\vec{k}^*=\left(l,-m\right)$. The frequencies are $\omega_{\vec{k}_1,\vec{k}_2}=l_1+l_2+1$ on $S^2$, so that the free energy of the singlet sector can be easily calculated as
\begin{align}
\begin{split}
F'_{\rm min}=&E^{(1)}\beta+\frac{1}{2}\sum_{\left(l_1,m_1\right),\left(l_2,m_2\right)}\log\left[1-e^{-\beta\left(l_1+l_2+1\right)}\right]+\frac{1}{2}\sum_{\left(l,m\right)}\log\left[1-e^{-\beta\left(2l+1\right)}\right]\\
=&E^{(1)}\beta-\frac{1}{2}\sum_{n=1}^\infty\frac{e^{-n\beta}}{n}\left[\frac{\left(1+e^{-n\beta}\right)^2}{\left(1-e^{-n\beta}\right)^4}+\frac{1+e^{-2n\beta}}{\left(1-e^{-2n\beta}\right)^2}\right]
\end{split}
\end{align}
where the factor $\frac{1}{2}$ in the second term are necessary for avoiding double-counting. Furthermore, the 1-loop correction to the ground state energy is $E^{\left(1\right)}=\frac{1}{2}\sum_{\vec{k}_1\le\vec{k}_2}\omega_{\vec{k}_1,\vec{k}_2}=\frac{1}{2}(K+1)\sum_{\vec{k}}\lambda(\vec{k})$ which precisely cancels the $\mathcal{O}(N^0)$ contribution from the counterterm $\Delta V$: $\Delta E^{(1)} = -\frac{1}{4} (K+1) \Tr \Psi_0^{-1}=-\frac{1}{2}(K+1)\sum_{\vec{k}}\lambda(\vec{k})$. This ensures the vanishing of the total one loop correction: $E^{\left(1\right)}_{\text{total}}=E^{\left(1\right)}+\Delta E^{\left(1\right)}=0$.
Therefore, the leftover correction to the free energy is
\begin{align}
\begin{split}
F_{\rm min}=&-\frac{1}{2}\sum_{n=1}^\infty\frac{e^{-n\beta}}{n}\left[\frac{\left(1+e^{-n\beta}\right)^2}{\left(1-e^{-n\beta}\right)^4}+\frac{1+e^{-2n\beta}}{\left(1-e^{-2n\beta}\right)^2}\right]
\end{split}
\end{align}
which agrees with \eqref{on} after the identification $q = e^{-\beta}$. 

In a similar way, one can calculate the free energy of singlet sector of $U\left(N\right)$ vector theory. In this case, the  bi-local field $\Psi\left(\vec{x},\vec{y}\right)$ is not symmetric hence there will be no potential double-counting, the final result is
\begin{align}
\begin{split}
F=&\sum_{\left(l_1,m_1\right),\left(l_2,m_2\right)}\log\left[1-e^{-\beta\left(l_1+l_2+1\right)}\right]=-\sum_{n=1}^\infty\frac{e^{-n\beta}}{n}\frac{\left(1+e^{-n\beta}\right)^2}{\left(1-e^{-n\beta}\right)^4}
\end{split}
\end{align}
which agrees with \eqref{un}. At high temperature, the free energy scales as $F \sim -4 \zeta\left(5\right) T^4$, showing the higher phase of \cite{Shenker:2011zf}.

What we have seen in the present series of calculations is that in this case the free energies do not vanish at one loop. But the ground state energy is indeed much like the free energy of the previous section: one obtains the exact result in the leading evaluation while the one loop contribution cancels with the contribution from the counterterm. The picture regarding the redefinition of the coupling constant in the $O(N)$ case therefore appears in this background too. That is satisfactory as there should not be a change in the identification of the coupling constant just by changing the background.

\section{Conclusions}
\label{sec:conclusions}

The purpose of the calculations presented in this paper is two-fold. First, the calculations (essentially of the determinants) in the bi-local collective picture and the AdS picture were seen to give identical results. This can serve as a further confirmation of the exact equivalence of the two pictures at the level of Laplacians representing small fluctuations. This agreement was seen for a variety of backgrounds. We note that the bi-local picture provides an explanation of the curious observation of \cite{Giombi:2013fka} that the evaluation of one-loop AdS determinants (after summing over all spins and zeta function regularization) gives a result identical to that of a single local field.

Second, the calculations performed offer further data for a specification of the dictionary between higher spin gravities and vector field theories (both $U(N)$ and $O(N)$). The issue (addressed in \cite{Giombi:2013fka}) is the possible difference between the identification of $G$ (the coupling constant in the higher spin theory) with the parameter $1/N$ or $1/(N-1)$ of the vector model. The collective field representation shows that the $1/N$ expansion can be maintained due to a presence of a measure factor in the functional integration. Due to the specific form of the measure we have also observed that for $O(N)$ based theories one can define an effective coupling constant given by $G'=1/(N-1)$ as proposed in \cite{Giombi:2013fka}. We caution however that this might be a gauge dependent phenomenon and might not be a feature of an arbitrary gauge. This problem (of measure)  is well understood in bi-local version of higher spin theory but remains to be understood in Vasiliev type gravities. For this, one requires the knowledge of the action \cite{Boulanger:2011dd}, which then through its (nonlinear) Poisson structure determines the measure appearing in the functional integral. Work on this is in progress. 

Finally let us mention that the collective approach can be extended to the interacting IR fixed point as well. For the AdS computations including the heat-kernel method and the Hamiltonian method, adapting to the IR fixed point is relatively straightforward just by changing the conformal dimension of the bulk scalar field from $\Delta_{(0)}=1$ to $\Delta_{(0)}=2$. On the field theory side, the IR fixed point can be reached either by turning on a double trace deformation or using the nonlinear sigma model. The collective approach can be applied to either method and we have checked that the free energy decreases along the RG flow (the F-theorem) with the difference $F_{IR}-F_{UV}=-\frac{\zeta(3)}{8 \pi^2}$ in the case of $S^3$ \cite{Klebanov:2011gs}. Most recently, an extension of the results in \cite{Giombi:2013fka} to other dimensions $d$ was performed in \cite{Giombi:2014iua}. It is clear that some of our conclusions from $d=3$ indeed hold more generally for other dimensions. Most important among them is the reduction of the small fluctuation determinant to a single local field one.

\acknowledgments

We would like to thank S. Das, M. Gaberdiel, S. Giombi, I. Klebanov, R. Leigh, A. Petkou and especially Sumit Das for helpful discussions. The work of AJ and JY is supported by the Department of Energy under contract DE-FG02-91ER40688. The work of KJ is supported by the DOE grant DE-FG02-13ER42001.

\appendix

\section{Derivation of the measure}
\label{app:measure}

In the transformation from the vector fields $\phi^i\left(x\right)$ to the bi-local fields $\Phi\left(x,y\right)$, the partition function gets a Jacobian
\begin{align}
\int \mathcal{D}\vec{\phi}\left(x\right)e^{-S\left[\vec{\phi}\right]}=\int \mathcal{D}\Phi\left(x,y\right) J(x,y) \, e^{-S\left[\Phi\left(x,y\right)\right]} \ .
\end{align}
For the symmetric bi-local field $\Phi\left(x,y\right)$, we can give an ordering to the spacetime coordinates $x$ and $y$.\footnote{For example, for $x=\left(x_1,x_2,\cdots,x_n\right)$ and $y=\left(y_1,y_2,\cdots,y_n\right)$, $x<y$ means iff $x_1<y_1$ or $x_1=y_1, x_2<y_2$ etc.} Then, the independent bi-local fields are
\begin{align}
\Phi\left(x,y\right)\qquad\left(x \le y\right) \ .
\end{align}
Denoting $a \equiv \left(x,y\right)$ and $b \equiv \left(x',y'\right)$ with $x \le y$ and $x' \le y'$, respectively, in general, one can derive a differential equation for the Jacobian \cite{Jevicki:1980zg}
\begin{equation}
\int dx'dy' \Omega\left(a,b\right)\frac{\partial \log J}{\partial \Phi\left(b\right)}+\omega\left(a\right)+\int dx'dy' \frac{\partial \Omega\left(a,b\right)}{\partial \Phi\left(b\right)}=0
\label{diff}
\end{equation}
where $\omega\left(a\right)$ and $\Omega\left(a,b\right)$ are $\mathcal{O}\left(N\right)$ and $\mathcal{O}\left(N^0\right)$ contributions to $\log J$, respectively
\begin{eqnarray}
\omega\left(a\right)&\equiv&-\int dz \frac{\partial^2}{\partial \phi^i\left(z\right)\partial \phi^i\left(z\right)}\Phi\left(a\right)=-2N\delta\left(x-y\right) \label{a4} \\
\Omega\left(a,b\right)&\equiv&\int dz\frac{\partial\Phi\left(a\right)}{\partial\phi^i\left(z\right)}\frac{\partial\Phi\left(b\right)}{\partial\phi^i\left(z\right)}\sim\mathcal{O}\left(N^0\right) \ . \label{a5}
\end{eqnarray}
One can also compute
\begin{align}
\int dx'dy' \frac{\partial \Omega\left(a,b\right)}{\partial \Phi\left(b\right)}=4\kappa \, \delta\left(x-y\right)
\label{a6}
\end{align}
where the coefficient $\kappa$ depends on the type of bi-local collective field theory
\begin{align}
\kappa=\begin{cases}
K&\qquad \text{for}~ U\left(N\right)\\
\frac{1}{2}\left(K+1\right)&\qquad \text{for}~ O\left(N\right)\\
\end{cases}
\end{align}
and $K \equiv \int dx \, \delta (0)=\sum_{k} 1$. 

Plugging (\ref{a4}) and (\ref{a6}) into the differential equation (\ref{diff}), we get
\begin{equation}
\int dx'dy' \Omega\left(a,b\right)\frac{\partial \log J}{\partial \Phi\left(b\right)}-2\left(N-2\kappa\right)\delta\left(x-y\right)=0 \ ,
\end{equation}
from which one can solve for the Jacobian for general collective field theory (see also \cite{Rodrigues:1992ru}) as
\begin{align}
\log J=\frac{1}{2}\left(N-2\kappa\right)\Tr\log \Phi \ ,
\end{align}
where we have used the explicit form of (\ref{a5}).

The parameter $\kappa$ is also related to weight of bi-local space when we express a function of bi-local space in terms of a function of local space. For example,
\begin{align}
\sum_{k_1, k_2} {}' \left[\lambda\left(k_1\right)+\lambda\left(k_2\right)\right]=2\kappa\sum_k \lambda\left(k\right)
\end{align}
where $\sum'$ means the summation over the independent bi-local momentum. For the $O(N)$ vector model, we have $\sum'_{k_1,k_2}=\sum_{k \le k_2}$ and $\kappa=\frac{1}{2}(K+1)$. While for the $U(N)$ vector model, there is no restriction on the summation, therefore $\kappa=K$.


\end{document}